# Phonon thermal transport properties of GaN with symmetry-breaking and lattice deformation induced by the electric field

Dao-Sheng Tang, Bing-Yang Cao*

(Key laboratory of thermal science and power engineering of Education of Ministry,

Department of Engineering Mechanics, Tsinghua University, Beijing 100084, China)

*Corresponding author, Email: caoby@tsinghua.edu.cn

**ABSTRACT:**

Electric fields commonly exist in semiconductor structures of electronics, bringing to bear on phonon thermal transport. Also, it is a popular method to tune thermal transport in solids. In this work, phonon and thermal transport properties of GaN with wurtzite and zincblende structures in the finite electric field are investigated using first-principles calculations from the perspectives of symmetry-breaking and lattice deformation. Effects of electric field on phonon transport properties including phonon dispersion and thermal conductivity from the response of electron density distribution only and response from lattice changes are studied in zincblende GaN. It is found that the former has a small but qualitative impact on phonon dispersion relations, i.e., splitting of phonon branches, since it breaks the symmetry of zincblende lattice. While the latter affects both lattice symmetry and size, causing significant changes in phonon properties and an increase in thermal conductivity. In wurtzite GaN, space-group-conserved lattice changes in the finite electric field are studied with lattice deformation only, where thermal conductivity decreases at electric fields significantly with the increase of anisotropy, much different from the changes in zincblende GaN. This work provides a comprehensive understanding of phonon thermal transport properties in GaN in the finite electric field, which promises to benefit phonon transport tuning and provide a reference for thermal management in GaN-based information and power electronics.

# I. Introduction

Generally, electric field exists in semiconductor structures of electronics since it plays a key role in manipulating electrons. Specifically, in GaN-based high electron mobility transistors (HEMTs), it is supposed that the high density of electrons at heterostructure results from strong polarization at AlGaN/GaN interface [1,2]. As one of the most important semiconductor materials, thermal transport in GaN has received great attention in both scientific researches and industrial applications [3-8]. However, the effects of electric fields on lattice thermal transport in GaN are still unclear, which is critical for thermal management in GaN-based transistors [4]. Besides, applying an external electric field is an effective way to tune the electrical and thermal transport properties of dielectric materials [9-15]. Theoretically, the response of lattice to the finite homogenous electric field can be classified as three parts [16,17]: (i) response of electron wave function (density distribution), which can be represented by Born effective charges and dielectric function, as well as interatomic force constants in phonon calculations; (ii) changes in atomic coordinates; and (iii) lattice strain, including electronic part and lattice part. Hence, for materials of which electronic or lattice response to electric field is strong, their thermal conductivity promises to be tuned by external electric field.

By performing first-principles calculations, several studies have been carried out for two-dimensional and layered materials with out-of-plane external electric fields. In detail, external electric field is used to tune electronic structures of two-dimensional BN, $C_2N$-h2D, graphene, and layered germanane [18-21], as well as phonon properties and lattice thermal conductivity of layered graphene [22], silicene [23], and borophene [24]. In Qin's work [23], the ultra-low thermal conductivity of silicene was obtained with external electric fields which induced phonon renormalization by affecting interatomic force constants. With similar mechanisms, anisotropy of thermal transport in borophene is manipulated by out-of-plane electric fields [24]. Besides affecting the magnitude and distribution of electronic charges, out-of-plane external electric fields can also break the inversion symmetry of layered structures, e.g., layered graphene, mixing in-plane optical phonon branches [22].

Compared with responses of two-dimensional materials, responses of three-dimensional periodic solids are much more abundant. External electric fields not only affect magnitudes of lattice constants and interatomic force constants, but also leave qualitative changes for lattice vibration and thermal transport by inducing structure phase transition and symmetry-breaking. However, investigations of thermal transport properties in three-dimensional periodic lattice under finite electric field are quite limited, since calculating the responses of three-dimensional periodic lattice structure to finite electric field from first principles is not an easy work. The main difficulty is that the scalar potential “$-\mathbf{E}\cdot\mathbf{r}$” ($\mathbf{E}$ is the electric field and $\mathbf{r}$ is the position vector) is nonperiodic and unbounded from below [25]. Under the framework of the modern theory of polarization, Souza et al. proposed an appropriate variance method based on the minimization of electric enthalpy functional [25]. Then, methods for calculating total energy of periodic solids as well as forces and stress, Born effective charges, dielectric function, and phonon properties from first principles were proposed and implemented in first-principles calculation software [16,17,26,27]. Later, an efficient approach to determine the optimized structure in finite electric field was developed based on the above understanding by modifying free energy and Hellmann-Feynman forces with Born effective charges and polarization, where the response of electronic part to electric field is ignored with approximations [26]. In cases where the response of electronic part can be ignored, this method has been applied well to study the lattice thermal transport accompanied by structure phase transition. As a result, it is now feasible to investigate the thermal transport properties of three-dimensional periodic solids [12-14,28]. Bagnall et al. [28] investigated the variance of optical phonons in wurtzite GaN induced by the changes in atomic coordinates from electric fields, and explained the shift of Raman peak positions accurately. In Liu's work [13], a larger thermal conductivity switch ratio in Barium Titanate was realized with external electric fields which can manipulate structure phase transition. Also, domain wall response to electric fields was introduced in ferroelectric materials, besides the responses mentioned above, which further enhances bidirectional tuning of thermal conductivity [12,14].

While the wurtzite structure GaN is thermodynamically stable and widely used in information and power electronics, the zincblende structure performs better in several other aspects, e.g., it is more suitable for light-emitting devices due to the absence of built-in piezoelectric fields and spontaneous polarization effects [29,30]. In this work, we perform first-principles calculations on GaN with these two lattice structures at external electric fields for two main purposes. One is to understand the response of lattice thermal transport of wurtzite GaN to external electric fields, and the other is to analyze different responses of lattice thermal transport to external electric fields in different structures of GaN, i.e., symmetry-breaking and symmetry-conserved cases. It is found that both positive and negative electric fields decrease the thermal conductivity of wurtzite GaN, and the responses of thermal transport in wurtzite and zincblende GaN to electric fields are significantly different.

## II. Methods

First-principles calculations in this work include phonon properties, e.g., phonon dispersion relations and density of states (DOS), and lattice thermal conductivity calculations. The phonon calculations are performed based on density functional theory as implemented in ABINIT [27,31] with optimized norm-conserving (ONCV) pseudopotential, while the lattice thermal conductivity calculations are based on Vienna ab initio Simulation Package (VASP) [32] with projected augmented wave (PAW) pseudopotential [33] since several tests show that ABINIT and Phono3py [34] based calculations provide much lower thermal conductivity of wurtzite GaN compared with the data from calculations and experiments in literature [3,5-7]. Generalized gradient approximation in the Perdew-Burke-Ernzerhof form [35] is adopted for the exchange-correlation functional. The kinetic energy cutoff for plane-wave basis 1000 eV is employed with strict convergence test, and the Brillouin zones are sampled using converged 8×8×6 Gamma-centered and 10×10×10 Monkhorst-Pack *k*-mesh grids [36] for wurtzite and zincblende structures, respectively. In the structural optimization step, the atom positions and lattice constants are fully relaxed until the residual stress and the maximum forces acting on each atom are smaller than $10^{-2}$ kbar and $10^{-6}$ eV/Å, respectively, for both cases with and without external electric field. The 2nd order interatomic force constants (IFCs) are calculated using a finite-displacement supercell method implemented in Phonopy [34] with supercell size 4×4×3 and 5×5×5 for wurtzite and zincblende primitive unit cell, respectively. The Born effective charges and high-frequency dielectric constants are calculated with density functional perturbation theory to include the long-range Coulomb atomic interaction in polar materials, i.e., nonanalytical corrections for 2nd order IFCs. The finite-displacement supercell method is also applied in the calculations of 3rd order IFCs using thirdorder.py, a script in ShengBTE package [37] with supercell size 3×3×3 and 4×4×4 for wurtzite and zincblende primitive unit cell, respectively. And the interaction cutoff is up to the fifth-nearest neighbors in the 3rd IFCs calculations. Lattice thermal conductivity is calculated by directly solving the phonon Boltzmann transport equation [37] with second and third interatomic force constants, where the size of $q$-mesh grids in the Brillouin zone are both 19×19×19 for wurtzite and zincblende structures. The parameters used in first-principles calculations as mentioned above have been carefully determined for convergence of the calculations based on literature [5-8] and our tests [3].

The case of three-dimensional solids in finite electric field corresponds to that the bulk material (or slab structure thick enough) is clamped by two electrodes with voltage difference in reality. In the relaxation process of bulk material, the absolute electric field changes with the variance of lattice constants while the voltage difference is generally constant. Hence, it is more appropriate to use the reduced electric field here which is defined as [38]

$$\bar{\varepsilon}_i = \mathbf{a}_i \cdot \boldsymbol{\varepsilon}, \qquad (1)$$

where $\mathbf{a}_i$ is the lattice vector and $\boldsymbol{\varepsilon}$ is the absolute electric field with their directions defined in Cartesian coordinates. In the following calculations of structural relaxation at finite electric field, the reduced electric field is used, instead of the absolute electric field.

In this work, forces on atoms, the total energy of the system, and lattice structure optimization in three-dimensional solids at finite electric field are calculated using open-source software ABINIT which can take into consideration the response from the electronic part, i.e., changes in electron density distribution. As mentioned above, since the combination of ABINIT and Phono3py gives a much lower thermal conductivity compared with data from literature including calculations and experiments with plenty of tests, the thermal conductivity of GaN is calculated based on the VASP package. Applying finite electric field in every supercell calculation in lattice thermal conductivity calculations is time-consuming with an unacceptable amount of computation, even for converged phonon dispersion calculations. Consequently, besides studies on the response of electron density distribution to external electric field particularly, the electric field is not applied in supercell calculations while only optimized lattice structures from the electric field are used. The rationality of this treatment lies in that electronic response in the finite electric field is very small quantitatively especially in space-group-conserved cases and the forces acting on ions from the electric fields can be eliminated in difference calculations for IFCs. As a summary for all calculations above, the formal calculations can be divided into three parts. The first part is phonon calculations of zincblende GaN at the electric field with fixed lattice and atoms. In this part, phonon dispersion relations are the main results, which are calculated by the finite-displacement supercell method as implemented in ABINIT and Phonopy packages. The second part is phonon and thermal conductivity calculations of zincblende GaN at different electric fields (0, positive, and negative electric fields). At first, optimized lattice structures at different electric fields are obtained by the ABINIT package. With these optimized lattice structures, phonon dispersion relations are calculated using the finite-displacement supercell method as implemented in ABINIT and Phonopy, and lattice thermal conductivities are calculated based on the IFCs from the finite-displacement supercell method and phonon Boltzmann transport equation using VASP, Phonopy, and ShengBTE. Supercell calculations are required in the finite-displacement supercell method for 2nd and 3rd IFCs, where the electric field is not implemented. The third part is phonon and thermal conductivity calculations of wurtzite GaN at different electric fields. Details in calculations are the same as those in the second part.

## III. Results and discussion

### A. Symmetry-breaking from the finite electric field

In this section, it is concentrated on the changes in phonon and phonon transport properties from symmetry-breaking of lattice system at the electric field, including symmetry-breaking from the electric field with fixed lattice and atomic positions, i.e., from changes in electron density distribution only, and electric field with lattice deformation, i.e., from lattice strain and changes in atomic positions at finite electric field.

For zincblende structure, its space group is F-43m with 24 lattice symmetry operations including 2-, 3-, and 4- fold rotation symmetry and mirror symmetry [39]. The detailed lattice structure and atomic positions are shown in Figure 1. Phonon dispersion relations contain six branches including three acoustic branches and three optical branches. The phonon degeneracies at high symmetry points are listed in Table 1, and they can also be seen from phonon dispersion relations. Since the three directions $x, y, z$ in Cartesian coordinates are equivalent for zincblende structures, reduced electric field along $z$-direction is applied (Figure 1). The breakdown electric fields ($5\times10^8$ V/m for zincblende GaN and $3.3\times10^8$ V/m for wurtzite GaN) from Chow's work [40] in 1996 are generally used. These values provide reference to experimental studies for increasing breakdown electric fields, while larger values may be used in numerical and theoretical studies in an ideal GaN system, e.g., electric field $5\times10^8$ V/m is used for wurtzite GaN in Ref. [28]. In this study for effects of symmetry-breaking and lattice deformation induced by the electric field on phonon transport properties, we intend to apply electric field as large as possible to observe the significant response of phonon transport properties to electric fields, which can also provide information on the feasibility of thermal conductivity tuning based on electric fields. In first-principles calculations at finite electric fields, only valence band electrons are considered, which regards the system as an insulator with infinite bandgap and implies that no limitation exists for electric fields. The experimental bandgap values are used as a criterion in this work to help select high electric fields to ensure reasonability and significant responses (an estimated lower limit for bandgap can be provided to a specific electric field in ABINIT). Limited by bandgap (3.23 eV) of zincblende GaN [41], absolute electric fields used in studies on electronic response are set to be 0.001 and 0.002 in atomic unit corresponding to $5.14\times10^{8}$ V/m and $1.03\times10^{9}$ V/m, and reduced electric fields in structural relaxation calculations are selected to be -0.011 and 0.011 in the atomic unit, corresponding to the absolute electric field $-1.316\times10^{9}$ V/m and $1.315\times10^{9}$ V/m, respectively.

In calculations with fixed lattice structure and atomic positions, only the changes in electron density distribution are considered. In this case, the symmetries of the lattice system are reduced, e.g., 2-fold rotation symmetry along $x$ ($y$) axis and mirror symmetry relying on the plane not containing $z$-axis do not exist and equivalence among $x$, $y$, $z$ directions is absent due to the unsymmetrical electron density distribution at finite electric field along the $z$-direction. It is noted here particularly that lattice symmetry is not used in phonon calculations to decrease computation costs as actual symmetry is lower than apparent lattice symmetry at finite electric fields. In Figure 2, phonon dispersion relations with and without finite electric field are presented. The electric field here is an absolute electric field, instead of the reduced electric field, since lattice structure and atomic positions are fixed during calculations. Static calculations of supercells in the finite electric field are very time-consuming, especially when the supercell is large. Hence, a 2×2×2 supercell size is used in phonon dispersion relation calculations in the finite electric field. Though the rough calculation may not provide exact quantitative conclusions, it can give a qualitative understanding. Since lattice structure and atomic positions are fixed and only electronic response is considered, there is almost no difference among them seen from full phonon dispersion relations along high symmetry paths in Figure 2 (a), which is understandable that quantitative effects of electronic response are very small. However, electronic response brings qualitative changes to phonon dispersion relations by introducing symmetry-breaking. At three high symmetry points Γ, X, and L where transverse optical (TO) phonon branches degenerate in the original zincblende structure, the degeneracies reduce due to the symmetry-breaking from the electric field, which is illustrated in Figures 2 (b)-(d). The gap between TO branches is relatively small in Figure 2(c) when $E_z$=0.001, which is not significantly distinguished from numerical errors, while the gaps are large enough in Figures 2(b) and (d). Here, two cases with $E_z$=0.001 and $E_z$=0.002 are studied to eliminate the inference from numerical errors, as a larger electric field increases the gap in magnitude compared with the smaller one.

With lattice response to the electric field being taken into consideration, lattice symmetry will be significantly reduced since lattice symmetry of the zincblende structure is sensitive to lattice deformation. It is noted here that applying positive and negative electric fields along *z*-direction are not equivalent. For zincblende GaN, there are two atoms including one Ga and one N atom in each unit cell. Due to the different atom types in the two sites, inversion symmetry does not exist in zincblende structures, and there is no mirror symmetry to the *x*-*y* plane. As a result, positive and negative directions along the *z*-axis in the zincblende structure are not equivalent. For a film or slab with a zincblende structure, the two surfaces are different, while the top surface is the N surface and the bottom surface is the Ga surface. The difference is also found in relaxed structures at electric fields, where internal distance in *z*-direction between Ga and N atoms inside the unit cell is decreased by the positive electric field and increased by the negative field. With considering lattice deformation only, i.e., ignoring electronic response, the space group of zincblende GaN changes from F-43m (No. 216) to R3m (No. 160) where only six symmetry operations exist, besides quantitative changes in lattice constant shown in Table 2 (See supplementary material for more details). Due to the symmetry-breaking from the electric field, phonon degeneracy is reduced in the Brillouin zone, which is obvious at high symmetry points as shown in Figures 3(a) and (b). The caption "free" in Figures 4, 5, and 6 means conditions with zero electric field. And the captions "Positive *E*" and "negative *E*" represent cases with reduced electric fields 0.011 and -0.011, respectively. To make comparisons better, the same paths in Brillouin zones are used in phonon dispersion plotting for original structure and symmetry-breaking structure. The changes mainly take place in optical phonon branches. With the reduction of phonon degeneracy, new phonon crossing points, which are accidental degenerate, are also generated in optical phonon branches at a positive electric field, shown in Figure 3(b). The reduction of phonon degeneracy and the generation of new accidental degenerate phonon crossing points from symmetry-breaking may lead to changes in topological properties of phonon systems [42], which is a potential application of electric fields. The DOS shown in Figure 3(c) are not significantly affected by the electric field as changes in them are relatively small. It is known that the change in harmonic force constants is the direct reason for the changes in phonon dispersions. In the calculations at electric fields with fixed lattice constants and atomic positions, changes in phonon dispersions result from the electronic response reflected by changes in both Born effective charges which induce the changes in harmonic force constants. Actually, changes in Born effective charges at finite electric fields from electronic and lattice responses are very small. In the calculations at finite electric fields with lattice deformation, changes in phonon dispersions result from lattice response, which is mainly reflected by the changes in harmonic force constants from lattice deformation while the electronic response is supposed to be small and not included in the calculations.

Besides phonon properties, phonon transport properties with symmetry-breaking from lattice deformation in the finite electric field are also discussed here. The relaxed structures of zincblende GaN at zero electric field between two packages (ABINIT and VASP) show good agreement, and the discrepancy is much smaller than the lattice structure changes from the larger external electric field, confirming the rationality to use VASP in lattice thermal conductivity calculations with relaxed structures in the finite electric field from ABINIT for zincblende GaN (See supplementary material for more details of phonon dispersions). Thermal conductivities of zincblende GaN with and without electric field are illustrated in Figure 4, as well as cumulative thermal conductivities with respect to phonon mean free path (MFP) and frequency in Figure 5. Since the electric field breaks lattice symmetry by lattice deformation, lattice thermal conductivity changes both quantitatively and qualitatively. Without external electric field, non-diagonal elements of thermal conductivity tensor are zero and the thermal conductivity is isotropic, as thermal conductivity tensor is

$$\kappa = \begin{bmatrix} 181.25 & 0 & 0 \\ 0 & 181.25 & 0 \\ 0 & 0 & 181.25 \end{bmatrix}. \tag{2}$$

When external electric fields are applied, thermal conductivity tensors in case of the positive electric field become

$$\kappa = \begin{bmatrix} 232.47 & -3.97 & 3.97 \\ -3.97 & 232.47 & 3.97 \\ 3.97 & 3.97 & 232.47 \end{bmatrix}. \tag{3}$$

Those in case of the negative electric field become

$$\kappa = \begin{bmatrix} 187.20 & 4.58 & -4.58 \\ 4.58 & 187.20 & -4.58 \\ -4.58 & -4.58 & 187.20 \end{bmatrix}. \tag{4}$$

Due to the symmetry-breaking reflected by lattice structure and atomic positions, non-diagonal elements are not equal to zero, though they are not large compared to diagonal elements. Lattice with space group R3m belongs to the trigonal system and can be described by two approaches. The first approach is introduced here for comparison with the cubic zincblende structure. In this approach, lattice parameters $a=b=c$, and three solid angles $\alpha=\beta=\gamma<120°$, $\neq 90°$. A typical lattice with space group R3m has six atoms in each unit cell while only two atoms in each unit cell in zincblende GaN in the finite electric field in the actual calculations. Thus, it is more appropriate to state that zincblende GaN in the finite electric field has the same lattice symmetry operations as those in lattice with R3m. Seen from the optimized lattice structure and atomic positions of zincblende GaN in the finite electric field, the changes in lattice symmetry mainly result from the difference among three lattice vectors and changes in atomic positions (See supplementary material). Off-diagonal terms in thermal conductivity tensor are physically consistent and not uncommon for lattices with low symmetry, e.g., $\beta$-$Ga_2O_3$ in our previous calculations [43]. The off-diagonal term $k_{xy}$ is not equal to zero in $\beta$-$Ga_2O_3$ with a monoclinic lattice system. Basically, thermal conductivity is a tensor instead of a scalar. Terms in the tensor can be simplified with the aid of lattice symmetry for crystals with high lattice symmetry only, e.g., only diagonal terms exist for hexagonal lattice and only one independent term exists for cubic lattice. Indeed, non-zero off-diagonal terms in thermal conductivity tensor imply that directions of temperature gradient and heat flux are not consistent. Based on our understanding, this phenomenon results from deviation of the lattice to orthorhombic lattice, i.e., this is a consequence of lattice symmetry reduction. The negative components occur in off-diagonal terms, such as $k_{xy}$ and $k_{xz}$. For example, $k_{xy}$ means that a temperature gradient in the $x$-direction will induce a heat flux in the $y$-direction. And the negative value (or positive value) represents the direction of the heat flux in the $y$-direction. Despite that lattice symmetry is reduced by electric field which is supposed to reduce lattice thermal conductivity in general, lattice thermal conductivity increases at electric fields. Specifically, the negative electric field gives a slight increase while the positive field gives a significant increase to thermal conductivity. Figures 5(a) and (b) show the normalized cumulative thermal conductivity with respect to phonon MFP and frequency, respectively. With electric fields, larger maximum MFPs are verified as thermal conductivity keeps increasing until phonon MFP is up to around 4 $\mu$m at the positive electric field and 2 $\mu$m at the negative electric field while the original one is about 1 $\mu$m. Since there are large gaps between low and high-frequency phonon branches with and without electric field, and high-frequency phonon branches are relatively flat as shown in Figure 3, the contribution from high-frequency phonons is nearly zero as illustrated in Figure 5(b).

For a detailed understanding on changes in lattice thermal conductivity, phonon scattering rate, group velocity, and specific heat are calculated and plotted in Figure 6. Based on phonon Boltzmann transport equation, lattice thermal conductivity formula is derived as [37]

$$\kappa^{xx} = \frac{1}{k_{\mathrm{B}} T^2 \Omega N} \sum_{qv} f_0 (f_0 + 1) \left( \hbar \omega_{qv} \right)^2 v_{qv}^x F_{qv}^x. \tag{5}$$

The parameters $k_{\mathrm{B},}\ T, \Omega,$ and $N$ in the denominator are Boltzmann constant, temperature, the volume of the unit cell, and the number of $q$-points in calculations. $f_0$ is Bose-Einstein distribution function, $\omega_{qv}$ is phonon frequency for phonon with wave vector $q$ and branch $v$, $v_{qv}^x$ is $x$ component of phonon group velocity, and $F_{qv}^x$ can be regarded as effective phonon mean free path defined as the product of $x$ component of phonon group velocity and effective relaxation time $\tau_{qv}$. This equation can also be written as

$$\kappa^{xx} = \sum_{qv} c_{\mathrm{V},qv} v_{qv}^x v_{qv}^x \tau_{qv}, \tag{6}$$

with mode-specific heat $c_{\mathrm{V},qv}$. Scattering rate is a very important an-harmonic phonon property, depending on phonon dispersion relations and anharmonicity of atomic interaction. In Figure 6(a), it is found that the distribution of scattering rate at zero electric field is more concentrated. The differences in magnitude are not significant. The scattering rate at the negative electric field is relatively large while that at the positive electric field is relatively small particularly at frequency 0-5 THz from which around 40% thermal conductivity is contributed. The weighted phase space for three phonon processes [44] and Grüneisen parameters to phonon frequency for zincblende GaN at different electric fields are plotted in Figures 6(b) and (c). It is found that electric fields increase the weighted space group slightly which contributes to the increase of phonon scattering, while they decrease the phonon anharmonicity significantly. The contributions to the changes in thermal conductivity from the changes in phase space and phonon anharmonicity are opposite. As the comprehensive consequence of weighted space group and phonon anharmonicity, phonon scattering rates increase at electric fields for some phonon modes while they decrease for the other modes. The square of the *x* component of group velocity at the electric field is much larger than that without an electric field, indicating larger thermal conductivity, seen in Figure 6(d). Typically, the group velocity changes are accompanied by the changes in phonon dispersions. Here, the same paths, i.e., high symmetry paths in the original Brillouin zone of zincblende GaN, are used in phonon dispersion relations for zincblende GaN at different electric fields, which can provide consistent comparisons and shows reduction of phonon degeneracy clearly. However, since Brillouin zones are different from the original one and these paths are not high symmetry paths in the Brillouin zone of zincblende GaN with symmetry-breaking, phonon dispersion relations along these paths cannot fully reflect the phonon frequency distribution in the Brillouin zone.

Hence, phonon group velocities can be significantly different for conditions with different electric fields though phonon dispersion relations in the low-frequency section are similar. Nearly the same specific heat is illustrated in Figure 6(e), which is not supposed to result in changes in lattice thermal conductivity.

### B. Lattice deformation from the finite electric field

For wurtzite GaN, convergence for electronic calculations is difficult at the finite electric field perpendicular to the polar axis, as electric field along this direction breaks lattice symmetry, verifying robust lattice symmetry at finite electric field. In practical applications, it is more common that an electric field is applied along the polar axis (Figure 7). Therefore, space-group-conserved cases with electric fields along the polar axis are concentrated in this work, where the main response of lattice to external electric fields is lattice deformation. Considering the limit from bandgap (3.49 eV) of wurtzite GaN [45], reduced electric fields are selected to range from -0.02 and 0.02 in the atomic unit, corresponding to absolute electric fields $-1.035\times10^{9}$ V/m and $1.040\times10^{9}$ V/m, respectively. The relaxed structures of wurtzite GaN at the finite electric fields are shown in Tables 3 (See supplementary material for more details). The results at zero electric field between two packages (ABINIT and VASP) show good agreement, and the discrepancy is much smaller than the lattice constants changes from the external electric field, especially when reduced electric fields are equal to -0.02 and 0.02, confirming the rationality to use VASP in lattice thermal conductivity calculations with relaxed structures at finite electric field from ABINIT for wurtzite GaN (See supplementary material for more details of phonon dispersions).

Phonon and phonon thermal transport properties are shown in Figures 8-12. Positive *E* and negative *E* in these figures represent the cases with reduced electric fields 0.02 and -0.02, respectively. Similar to the discussion on zincblende GaN, phonon dispersion relations with and without lattice deformation from electric fields are illustrated separately for clear comparisons in Figures 8(a) and (b). Variances of acoustic branches and low-frequency optical branches along high symmetry paths are very small while significant changes take place in high-frequency optical branches, as well as DOS shown in Figure 8(c). Since lattice symmetry is conserved at finite electric fields, phonon degeneracies at high symmetry paths do not show obvious changes besides accidental degeneracy induced by lattice deformation. Different from the overall frequency shift and reduction with biaxial strain [3], the main changes here are shape changes while the overall frequency variances are not obvious. Also, the gap between high-frequency optical phonons and lower frequency phonons decreases at the finite electric field, which will increase phonon scattering generally. The optical phonon shifts in electric fields are consistent with those in Ref. [28] (See more detail in supplementary material).

In Figure 9(a), thermal conductivity with respect to a temperature ranging from 300-500 K is presented. Both positive and negative electric fields result in large decreases of thermal conductivity, opposite to the lattice thermal conductivity response to the electric field in zincblende GaN. Besides, anisotropy increases in the finite electric field while thermal conductivity is nearly isotropic at zero electric field. Here, thermal conductivity along the polar axis is regarded as out-of-plane thermal conductivity, and that in the direction perpendicular to the polar axis is in-plane one. In detail, the decrease of in-plane thermal conductivity at the positive electric field is relatively small while out-of-plane thermal conductivity decreases a lot. At negative electric field states, both in-plane and out-of-plane thermal conductivities decrease significantly, with room temperature data illustrated in Figure 9(b). Figures 9(c) and (d) show thermal conductivity contribution from each phonon band (bands 1-6) for in-plane and out-of-plane cases at room temperature. It can be seen from these data that thermal conductivity changes at electric fields mainly result from the contribution variation of phonons with higher frequencies in low-frequency parts. As it is difficult to distinguish phonon branches at non-Gamma points in the Brillouin zone, phonon branches here are distinguished according to the magnitude of phonon frequency (in ShengBTE) and are nonequivalent to the actual phonon branches. Figures 10(a) and (b) show the normalized cumulative thermal conductivity concerning phonon mean free path and frequency. As seen from the data in Figure 10(a), maximum MFPs at free state, i.e., at zero electric field, are around 10 $\mu$m. Corresponding to the variances of lattice thermal conductivity at finite electric fields, changes in maximum phonon MFPs are, however, not large. Maximum phonon MFPs nearly keep constant for the out-of-plane condition while a slight increase occurs for the in-plane condition. Since electric fields do not significantly shift or reduce low-frequency phonon branches, maximum cut-off frequencies in Figure 10(b) nearly keeps constant at electric fields.

Though significant changes take place for lattice thermal conductivity at finite electric fields, no obvious difference is found among phonon properties at zero, positive, and negative electric fields, which can explain the changes in lattice thermal conductivity. At the mode level, the absolute value of thermal conductivity at room temperature depends on phonon dispersion relation or specific heat, group velocity, and relaxation time (or its inverse, scattering rate). And its anisotropy depends on phonon group velocity particularly. Phonon scattering rate, *x*, *y*, and *z* components of group velocity, Grüneisen parameter, and specific heat are shown in Figures 11(a)-(g). In Figure 11(a), slightly larger scattering rates are found at electric fields. Comparing the square of the *x*(*y*/*z*) component of group velocity individually, it is difficult to conclude qualitatively from Figures 11(b)-(d). Square of the *x* component of group velocity is much different from that of the *y* component of group velocity, but the thermal conductivities in *x* and *y* directions are the same, illustrating that distribution of mode phonon properties plays an important role in determining lattice thermal conductivity, which is supposed to be responsible for changes in thermal conductivity at electric fields. At frequency 5-7 THz, the scattering rate increases in both positive and negative electric fields, especially in negative cases. As only relaxed structures in electric fields are used in phonon and thermal conductivity calculations, i.e., electric fields are not applied in supercell calculations, changes in phonon properties are attributed to the lattice deformation including atomic position changes and lattice strain. The relaxed structures (see supplemental material) show that lattice constant *a* decreases and *c* increases in positive fields while both *a* and *c* increase in negative fields. Also, we plot the weighted phase space for three phonon processes and Grüneisen parameters for further discussion on the scattering rate in Figures 11(e) and (f). No significant variance is found for weighted phase space, especially at the lower frequency part, which is reasonable as lattice symmetry of the deformed structure is conserved and phonon dispersions at the lower frequencies nearly keep constant. At frequency 5-7 THz, Grüneisen parameters increase in both positive and negative electric fields, implying the increase of the anharmonicity of interatomic interactions. The increases are consistent with the changes in phonon scattering rates at electric fields. To provide a quantitative understanding on the contributions to lattice thermal conductivity changes from phonon scattering rate, group velocity, and specific heat, changes (absolute value) in lattice thermal conductivity at restricted conditions [3] are calculated and shown in Figure 12, where “Relaxation time,” “Group velocity,” and “Specific heat” are marked as restricted conditions. “Relaxation time” means that only the relaxation time at positive electric field state is used, while group velocity and specific heat under free state are used in thermal conductivity calculation at positive electric field state. The results show that changes in relaxation times (or scattering rates) contribute the most to variances of lattice thermal conductivity at electric fields for both positive and negative conditions, while contributions from the changes in group velocity and specific heat can be ignored.

In this work, the isotope effect is not taken into consideration. While this effect is reported to be significant in first-principles calculations, it is found to be small in experiments. In calculations, the isotope effect is treated based on Tamura's mass variation approximation [46] and the Matthiessen rule. Since this type of scattering is not affected by electric fields, it only decreases the thermal conductivity of GaN at different electric fields simultaneously and will not affect the conclusions in this work.

## IV. Concluding remarks

In conclusion, phonon and phonon transport properties of zincblende and wurtzite GaN in the finite electric field are investigated systematically from the perspectives of symmetry-breaking and lattice deformation using first-principles calculations. The results show that responses of phonon and phonon thermal transport with different structures to external electric fields are significantly different. For zincblende GaN, though the electronic response to the electric field is small, the reduction of the degeneracy of transverse optical phonon branches is obvious due to the symmetry-breaking from unsymmetrical electrons density distribution at electric fields. Further symmetry-breaking induced by lattice deformation significantly changes the phonon dispersion relations, especially optical branches. Lattice symmetry-breaking does not necessarily decrease thermal conductivity as it increases thermal conductivity remarkably in the current case, mainly resulting from the increase of group velocity. Space-group-conserved relations at the electric field are performed on wurtzite GaN with robust lattice symmetry. Lattice thermal conductivity decreases significantly at both positive and negative electric fields, with increasing anisotropy of thermal conductivity. Changes in the distribution of mode phonon properties are responsible for the decrease of thermal conductivity at electric fields since no significant difference is found for phonon properties as a whole. And quantitative analyses confirm that the change of relaxation time is the main reason for the changes in lattice thermal conductivity at electric fields, which results from the increase of the anharmonicity of interatomic interactions.

## Acknowledgment

This work was supported by the National Natural Science Foundation of China (Nos. 51825601 and U20A20301). The authors appreciate the valuable comments and suggestions from the reviewers.

## Supplementary material

See supplementary materials for the benchmark study on AlAs, more details about lattice parameters with and without electric fields, phonon dispersion relations based on ABINIT and VASP, discussion on room temperature thermal conductivity changes concerning electric fields, optical phonon shifts, and $Q$-mesh convergence in thermal conductivity calculations for zincblende GaN.

## Data availability

The data that support the findings of this study are available from the corresponding author upon reasonable request.

## References

[1] J.P. Ibbetson, P.T. Fini, K.D. Ness, S.P. DenBaars, J.S. Speck, U.K. Mishra, Polarization effects, surface states, and the source of electrons in AlGaN/GaN heterostructure field effect transistors, Applied Physics Letters, 77(2) (2000) 250-252.

[2] O. Ambacher, J. Smart, J.R. Shealy, N.G. Weimann, K. Chu, M. Murphy, W.J. Schaff, L.F. Eastman, R. Dimitrov, L. Wittmer, M. Stutzmann, W. Rieger, J. Hilsenbeck, Two-dimensional electron gases induced by spontaneous and piezoelectric polarization charges in N- and Ga-face AlGaN/GaN heterostructures, Journal of Applied Physics, 85(6) (1999) 3222-3233.

[3] D.-S. Tang, G.-Z. Qin, M. Hu, B.-Y. Cao, Thermal transport properties of GaN with biaxial strain and electron-phonon coupling, Journal of Applied Physics, 127(3) (2020) 035102.

[4] Y.-C. Hua, H.-L. Li, B.-Y. Cao, Thermal Spreading Resistance in Ballistic-Diffusive Regime for GaN HEMTs, IEEE Transactions on Electron Devices, 66(8) (2019) 3296-3301.

[5] L. Lindsay, D.A. Broido, T.L. Reinecke, Ab initio thermal transport in compound semiconductors, Physical Review B, 87(16) (2013) 165201.

[6] L. Lindsay, D.A. Broido, T.L. Reinecke, Thermal conductivity and large isotope effect in GaN from first principles, Physical Review Letters, 109(9) (2012) 095901.

[7] A. Togo, L. Chaput, I. Tanaka, Distributions of phonon lifetimes in Brillouin zones, Physical Review B, 91(9) (2015) 094306.

[8] J.-Y. Yang, G. Qin, M. Hu, Nontrivial contribution of Fröhlich electron-phonon interaction to lattice thermal conductivity of wurtzite GaN, Applied Physics Letters, 109(24) (2016) 242103.

[9] S. Wang, J. Yu, Tuning electronic properties of silicane layers by tensile strain and external electric field: A first-principles study, Thin Solid Films, 654 (2018) 107-115.

[10] Q. Liu, L. Li, Y. Li, Z. Gao, Z. Chen, J. Lu, Tuning Electronic Structure of Bilayer MoS2 by Vertical Electric Field: A First-Principles Investigation, The Journal of Physical Chemistry C, 116(40) (2012) 21556-21562.

[11] K.H. Khoo, M.S.C. Mazzoni, S.G. Louie, Tuning the electronic properties of boron nitride nanotubes with transverse electric fields: A giant dc Stark effect, Physical Review B, 69(20) (2004) 201401(R).

[12] C. Liu, Y. Chen, C. Dames, Electric-Field-Controlled Thermal Switch in Ferroelectric Materials Using First-Principles Calculations and Domain-Wall Engineering, Physical Review Applied, 11(4) (2019) 044002.

[13] C. Liu, V. Mishra, Y. Chen, C. Dames, Large Thermal Conductivity Switch Ratio in Barium Titanate Under Electric Field through First-Principles Calculation, Advanced Theory and Simulations, 1(12) (2018) 1800098.

[14] C. Liu, P. Lu, Z. Gu, J. Yang, Y. Chen, Bidirectional Tuning of Thermal Conductivity in Ferroelectric Materials Using E-Controlled Hysteresis Characteristic Property, The Journal of Physical Chemistry C, 124(48) (2020) 26144-26152.

[15] Z.T. Zhang, R.Y. Dong, D.S. Qiao, B.Y. Cao, Tuning the thermal conductivity of nanoparticle suspensions by electric field, Nanotechnology, 31(46) (2020) 465403.

[16] X. Wang, D. Vanderbilt, First-principles perturbative computation of dielectric and Born charge tensors in finite electric fields, Physical Review B, 75(11) (2007) 115116.

[17] X. Wang, D. Vanderbilt, First-principles perturbative computation of phonon properties of insulators in finite electric fields, Physical Review B, 74(5) (2006) 054304.

[18] R. Zhang, B. Li, J. Yang, Effects of stacking order, layer number and external electric field on electronic structures of few-layer C2N-h2D, Nanoscale, 7(33) (2015) 14062-14070.

[19] Q. Tang, J. Bao, Y. Li, Z. Zhou, Z. Chen, Tuning band gaps of BN nanosheets and nanoribbons via interfacial dihalogen bonding and external electric field, Nanoscale, 6(15) (2014) 8624-8634.

[20] Y. Li, Z. Chen, Tuning Electronic Properties of Germanane Layers by External Electric Field and Biaxial Tensile Strain: A Computational Study, The Journal of Physical Chemistry C, 118(2) (2014) 1148-1154.

[21] H. Raza, E.C. Kan, Armchair graphene nanoribbons: Electronic structure and electric-field modulation, Physical Review B, 77(24) (2008) 245434.

[22] R. Stein, D. Hughes, J.-A. Yan, Electric-field effects on the optical vibrations in AB-stacked bilayer graphene, Physical Review B, 87(10) (2013) 100301(R).
[23] G. Qin, Z. Qin, S.Y. Yue, Q.B. Yan, M. Hu, External electric field driving the ultra-low thermal conductivity of silicene, Nanoscale, 9(21) (2017) 7227-7234.
[24] Z. Yang, K. Yuan, J. Meng, M. Hu, Electric field tuned anisotropic to isotropic thermal transport transition in monolayer borophene without altering its atomic structure, Nanoscale, 12(37) (2020) 19178-19190.
[25] I. Souza, J. Iniguez, D. Vanderbilt, First-principles approach to insulators in finite electric fields, Physical Review Letters, 89(11) (2002) 117602.
[26] H. Fu, L. Bellaiche, First-principles determination of electromechanical responses of solids under finite electric fields, Physical Review Letters, 91(5) (2003) 057601.
[27] X. Gonze, F. Jollet, F. Abreu Araujo, et al., Recent developments in the ABINIT software package, Computer Physics Communications, 205 (2016) 106-131.
[28] Kevin R. Bagnall, Cyrus E. Dreyer, David Vanderbilt , E.N. Wang, Electric field dependence of optical phonon frequencies in wurtzite GaN observed in GaN high electron mobility transistors, Journal of Applied Physics, 120(6) (2016) 155104.
[29] M.J.I. Khan, Z. Kanwal, N. Usmani, P. Akhtar, S. Hussain, Exploring optical properties of Gd doped zincblende GaN for novel optoelectronic applications (A DFT+U study), Materials Research Express, 6(11) (2019) 115916.
[30] L.Y. Lee, Cubic zincblende gallium nitride for green-wavelength light-emitting diodes, Materials Science and Technology, 33(14) (2017) 1570-1583.
[31] X. Gonze, B. Amadon, P.M. Anglade, et al., ABINIT: First-principles approach to material and nanosystem properties, Computer Physics Communications, 180(12) (2009) 2582-2615.
[32] G. Kresse, J. Furthmuller, Efficient iterative schemes for ab initio total-energy calculations using a plane-wave basis set, Physical Review B, 54 (1996) 11169.
[33] G. Kresse, D. Joubert, From ultrasoft pseudopotentials to the projector augmented-wave method, Physics Review B, 59 (1999) 1758.
[34] A. Togo, I. Tanaka, First principles phonon calculations in materials science, Scripta Materialia, 108 (2015) 1-5.
[35] John P. Perdew, Kieron Burke, M. Ernzerhof, Generalized Gradient Approximation Made Simple, Physical Review Letters, 77 (1996) 3865.
[36] H.J. Monkhorst, J.D. Pack, Special points for Brillouin-zone integrations, Physical Review B, 13(12) (1976) 5188-5192.
[37] W. Li, J. Carrete, N. A. Katcho, N. Mingo, ShengBTE: A solver of the Boltzmann transport equation for phonons, Computer Physics Communications, 185(6) (2014) 1747-1758.
[38] M. Stengel, N.A. Spaldin, D. Vanderbilt, Electric displacement as the fundamental variable in electronic-structure calculations, Nature Physics, 5(4) (2009) 304-308.
[39] M.I. Aroyo, A. Kirov, C. Capillas, J.M. Perez-Mato, H. Wondratschek, Bilbao Crystallographic Server. II. Representations of crystallographic point groups and space groups, Acta Crystallogr A, 62(Pt 2) (2006) 115-128.
[40] Chow, T.P, Ghezzo. SiC power devices. in III-Nitride, SiC, and Diamond Materials for Electronic Devices. Eds. Gaskill D.K, Brandt C.D. and Nemanich R.J., Material Research Society Symposium Proceedings, Pittsburgh, PA. 423 (1996) 69-73.

[41] G. Ramirez-Flores, H. Navarro-Contreras, A. Lastras-Martinez, R.C. Powell, J.E. Greene, Temperature-dependent optical band gap of the metastable zinc-blende structure beta -GaN, Physical Review B, 50(12) (1994) 8433-8438.

[42] D.S. Tang, B.Y. Cao, Topological effects of phonons in GaN and AlGaN: A potential perspective for tuning phonon transport, Journal of Applied Physics, 129 (2021) 085102.

[43] Y.B. Liu, J.Y. Yang, G.M. Xin, L.H. Liu, G. Csányi, B.Y. Cao. Machine learning interatomic potential developed for molecular simulations on thermal properties of $\beta$-$Ga_2O_3$. The Journal of Chemical Physics, 153 (2020) 144501.

[44] W. Li, N. Mingo, Ultralow lattice thermal conductivity of the fully filled skutterudite $YbFe_4Sb_{12}$ due to the flat avoided-crossing filler modes, Physical Review B, 91 (2015) 144304.

[45] B. Monemar, Fundamental energy gap of GaN from photoluminescence excitation spectra, Physical Review B, 10(2) (1974) 676-681.

[46] S. Tamura, Isotope scattering of large-wave-vector phonons in GaAs and InSb: Deformation-dipole and overlap-shell models, Physical Review B, 30 (1984) 849-854.

Table 1. Phonon degeneracy at high symmetry points at zero electric field

| *k*-point | Degenerated phonon branches |
|---|---|
| Γ | (1,2,3) (4,5) (6) |
| X | (1,2) (3) (4,5) (6) |
| W | (1) (2) (3) (4) (5) (6) |
| K | (1) (2) (3) (4) (5) (6) |
| L | (1,2) (3) (4,5) (6) |
| U | (1) (2) (3) (4) (5) (6) |

*The numbers 1,2, 3, … are band indexes where the bands in the same bracket are degenerate.

Table 2. Lattice constants of zincblende GaN at finite electric field

| Reduced *E* | *a* (Å) |
|---|---|
| 0.011 | 4.55318 |
| 0.006 | 4.55163 |
| 0 | 4.55020 |
| 0 (VASP) | 4.55005 |
| -0.006 | 4.54921 |
| -0.011 | 4.54871 |

Table 3. Lattice constants of wurtzite GaN at finite electric field

| Reduced *E* | *a* (Å) | *a*/*c* | *u* |
|---|---|---|---|
| 0.02 | 3.22430 | 1.62305 | 0.38290 |
| 0.01 | 3.22178 | 1.62603 | 0.37968 |
| 0 | 3.21970 | 1.62883 | 0.37676 |
| 0 (VASP) | 3.21922 | 1.62934 | 0.37670 |
| -0.01 | 3.21799 | 1.63151 | 0.37405 |
| -0.02 | 3.21664 | 1.63401 | 0.37153 |

## Figure captions

**Figure 1.** The lattice structure of zincblende GaN. The letter *a* indicates the lattice constant, and the arrow represents the direction of the positive electric field.

**Figure 2.** Phonon dispersion relations of zincblende GaN (a) along high symmetry paths (b) at Γ point (c) at X point, and (d) at L point in the finite electric field with fixed lattice structure and atomic positions. The electric field $E_z$ in figures is the absolute electric field, instead of the reduced electric field. The atomic unit (a.u.) is used for the electric field where 1 a.u. is equal to $5.14 \times 10^{11}$ V/m.

**Figure 3.** Phonon dispersion relations and DOS of zincblende GaN with and without symmetry-breaking from lattice deformation in the finite electric field. The caption "free" in the figures means a condition with zero electric field. And the captions "positive *E*" and "negative *E*" in the figures represent cases with reduced electric fields 0.011 and -0.011, respectively.

**Figure 4.** Lattice thermal conductivity of zincblende GaN with and without symmetry-breaking from lattice deformation at the electric field (a) to temperature (b) at room temperature.

**Figure 5.** Normalized cumulative lattice thermal conductivity of zincblende GaN to (a) phonon MFP and (b) phonon frequency, with and without symmetry-breaking from lattice deformation at the electric field at room temperature.

**Figure 6.** Phonon and thermal properties of zincblende GaN at room temperature with and without symmetry-breaking from lattice deformation at the electric field: (a) scattering rate (b) square of the *x* component of group velocity (c) weighted space group for three phonon process (d) Grüneisen parameters (e) specific heat.

**Figure 7.** The lattice structure of wurtzite GaN with lattice constants *a* and *c*, and internal parameter *u*. The arrow represents the direction of the positive electric field.

**Figure 8.** Phonon dispersion relations and DOS of wurtzite GaN with and without lattice deformation from the electric field. The caption "free" in the figures means a condition with zero electric field. And the captions “positive *E*” and “negative *E*” in the figures represent cases with reduced electric fields 0.02 and -0.02, respectively.

**Figure 9.** Lattice thermal conductivity of wurtzite GaN with and without lattice deformation from electric field (a) to temperature (b) at room temperature. And lattice thermal conductivity at room temperature for each low-frequency phonon band (bands 1-6) with and without the electric field for (c) in-plane and (d) out-of-plane cases.

**Figure 10.** Normalized cumulative lattice thermal conductivity of wurtzite GaN to (a) phonon MFP and (b) phonon frequency, with and without lattice deformation from the electric field at room temperature.

**Figure 11.** Phonon and thermal properties of wurtzite GaN at room temperature with and without lattice deformation from electric field (a) scattering rate (b) square of the $x$ component of group velocity (c) square of the $y$ component of group velocity (d) square of the $z$ component of group velocity (e) Grüneisen parameter (f) specific heat.

**Figure 12.** Changes in lattice thermal conductivity (absolute values) at restricted conditions with (a) positive electric field (b) negative electric field.

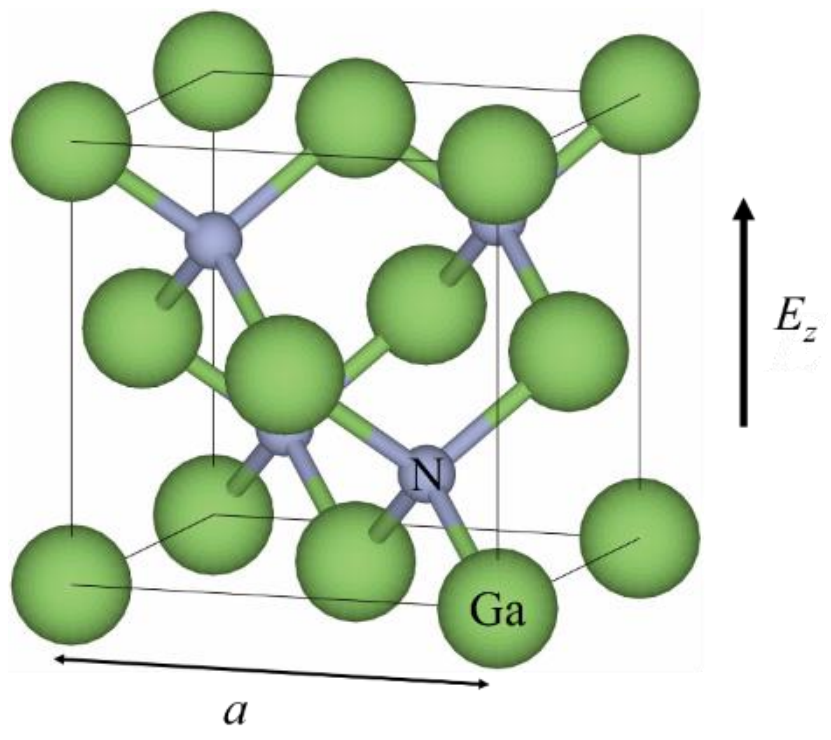

Figure 1. The lattice structure of zincblende GaN. The letter *a* indicates the lattice constant, and the arrow represents the direction of the positive electric field.

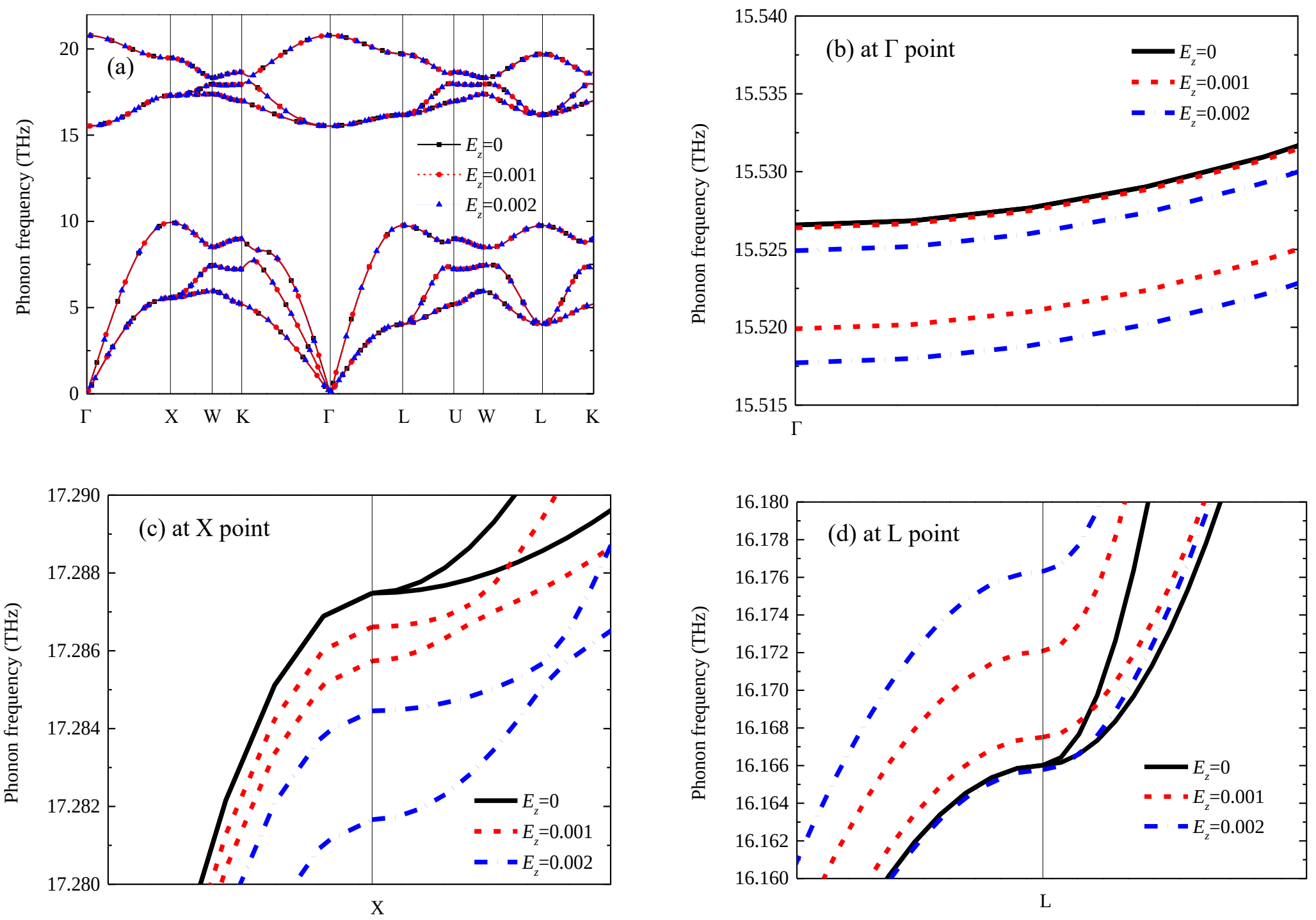

Figure 2. Phonon dispersion relations of zincblende GaN (a) along high symmetry paths (b) at Γ point (c) at X point, and (d) at L point in the finite electric field with fixed lattice structure and atomic positions. The electric field $E_z$ in figures is the absolute electric field, instead of the reduced electric field. The atomic unit (a.u.) is used for the electric field where 1 a.u. is equal to $5.14\times10^{11}$ V/m.

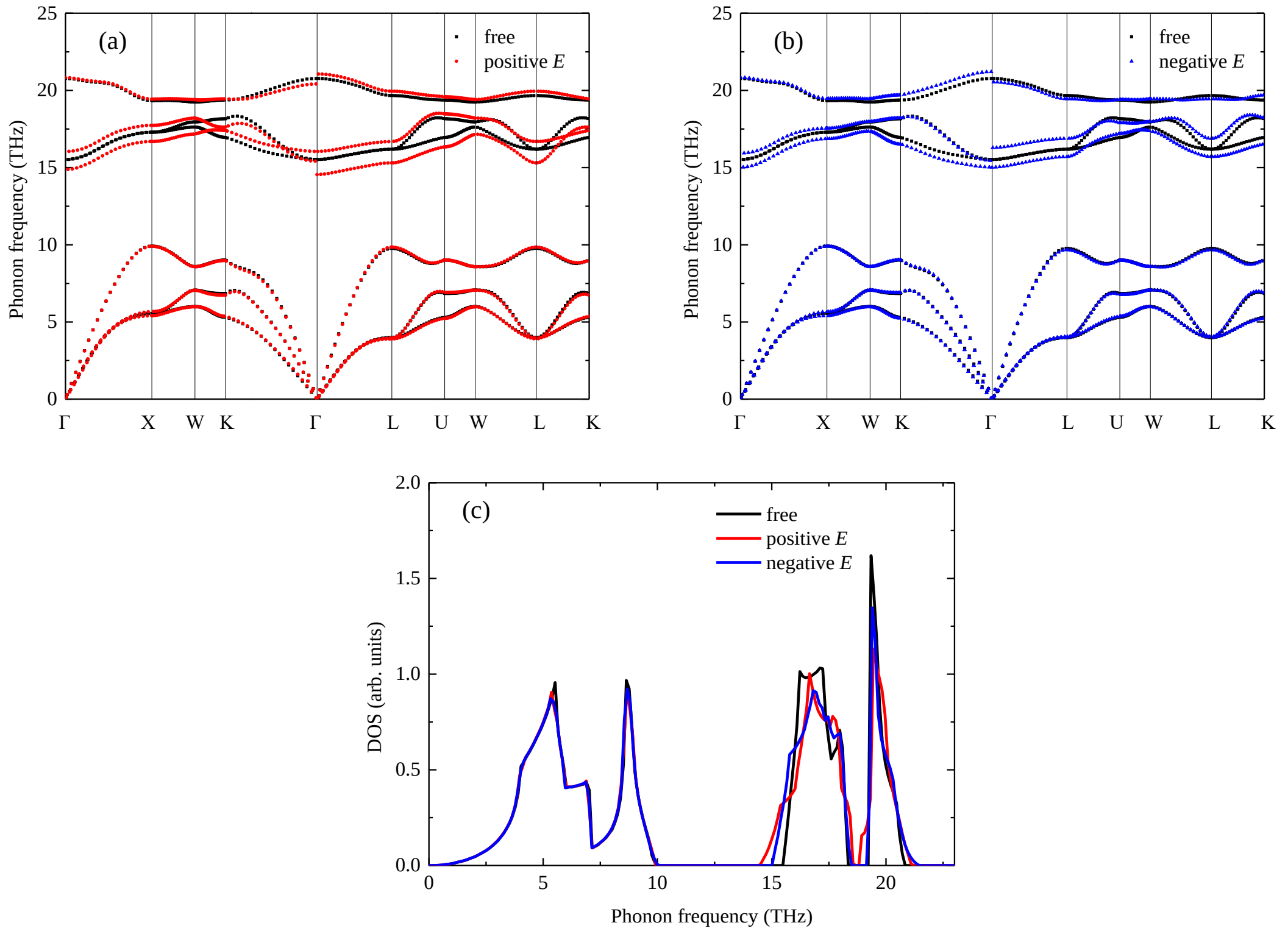

Figure 3. Phonon dispersion relations and DOS of zincblende GaN with and without symmetry-breaking from lattice deformation in the finite electric field. The caption "free" in the figures means a condition with zero electric field. And the captions "positive *E*" and "negative *E*" in the figures represent cases with reduced electric fields 0.011 and -0.011, respectively.

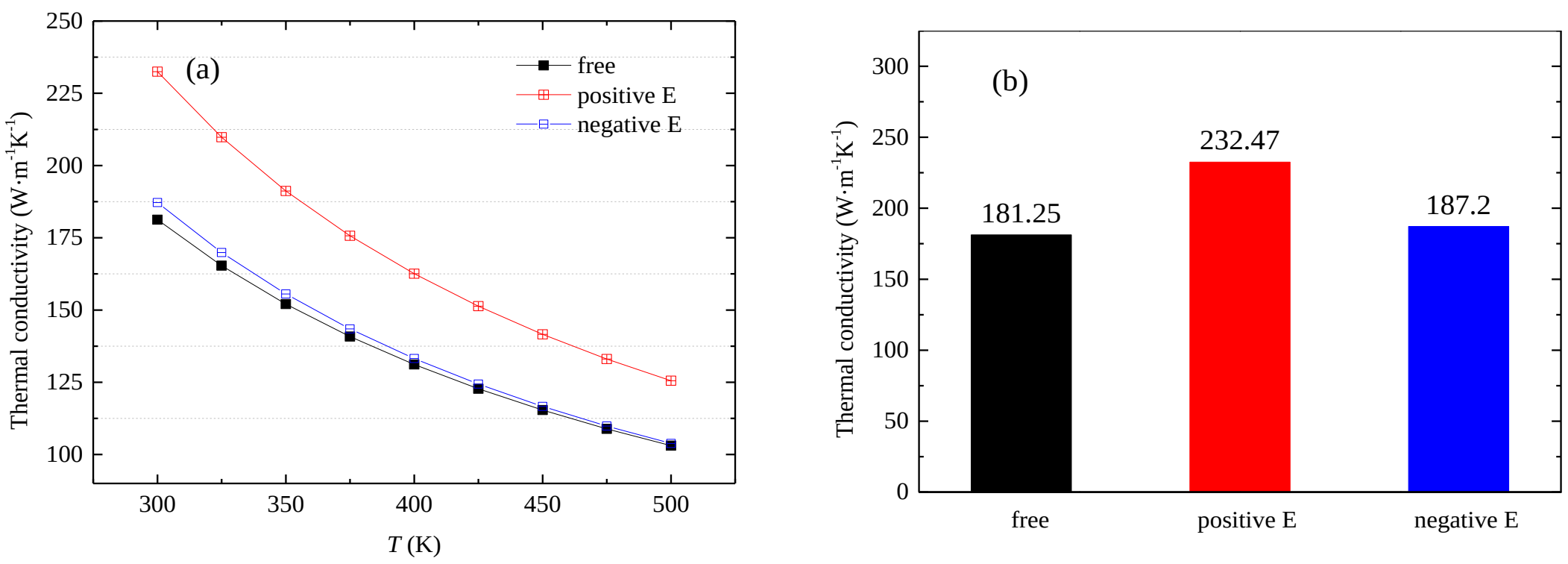

Figure 4. Lattice thermal conductivity of zincblende GaN with and without symmetry-breaking from lattice deformation at the electric field (a) to temperature (b) at room temperature.

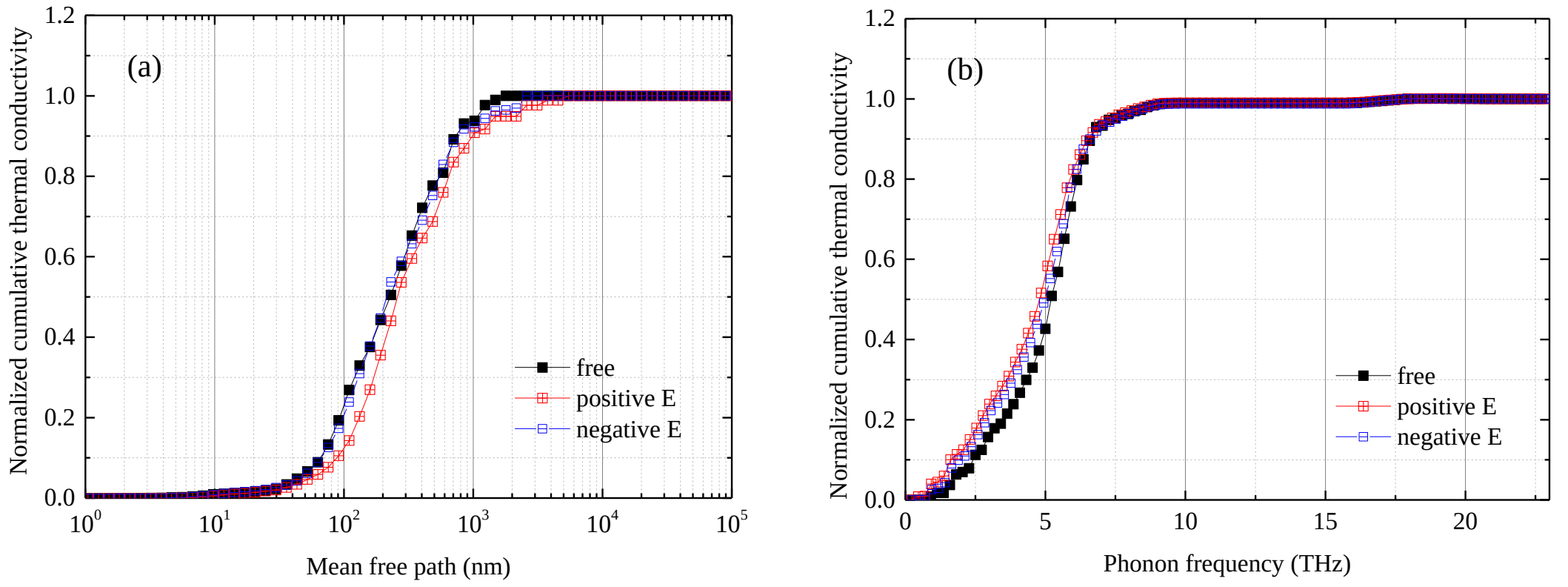

Figure 5. Normalized cumulative lattice thermal conductivity of zincblende GaN to (a) phonon MFP and (b) phonon frequency, with and without symmetry-breaking from lattice deformation at the electric field at room temperature.

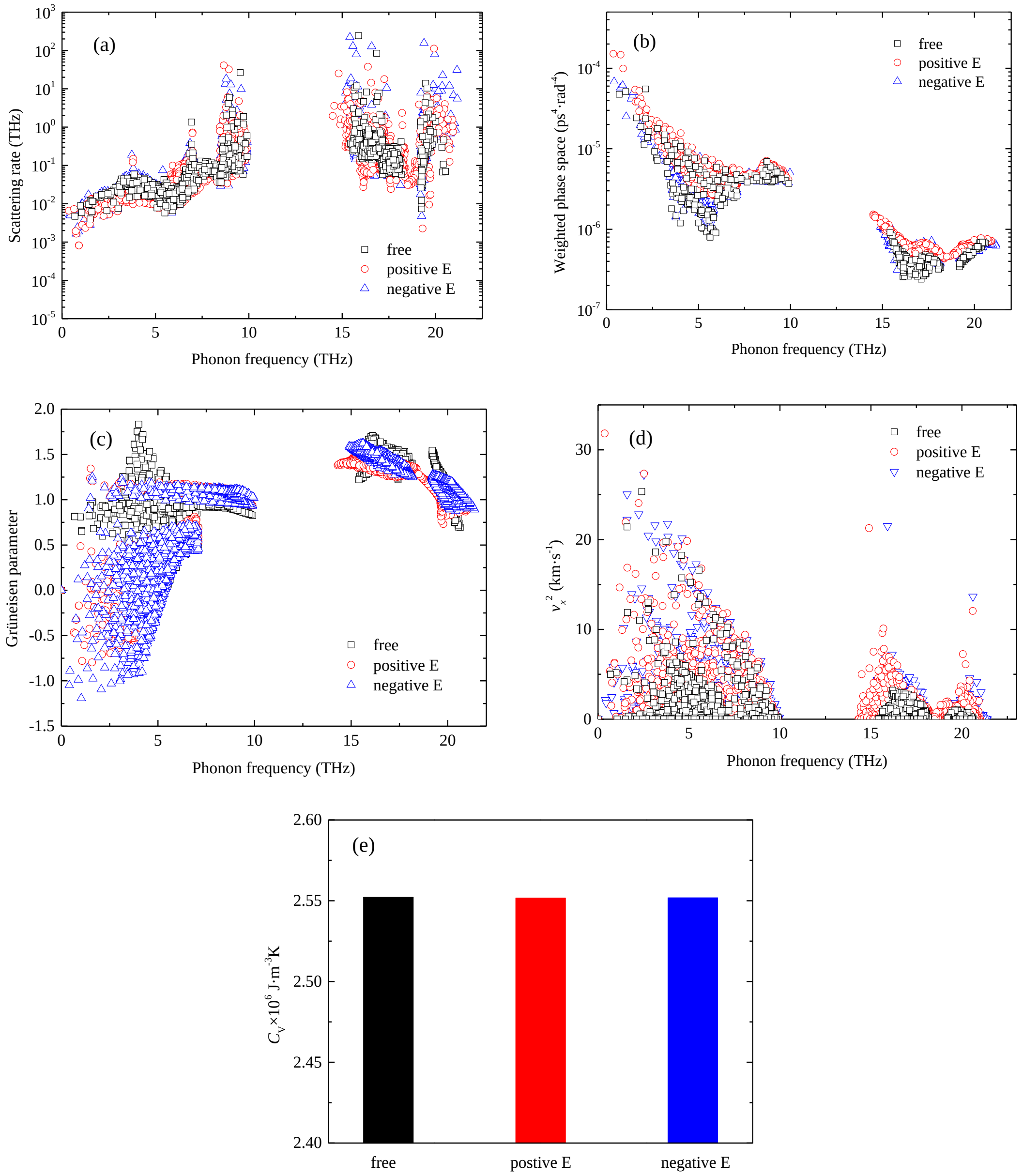

Figure 6. Phonon and thermal properties of zincblende GaN at room temperature with and without symmetry-breaking from lattice deformation at the electric field: (a) scattering rate (b) square of the *x* component of group velocity (c) weighted space group for three phonon process (d) Grüneisen parameters (e) specific heat.

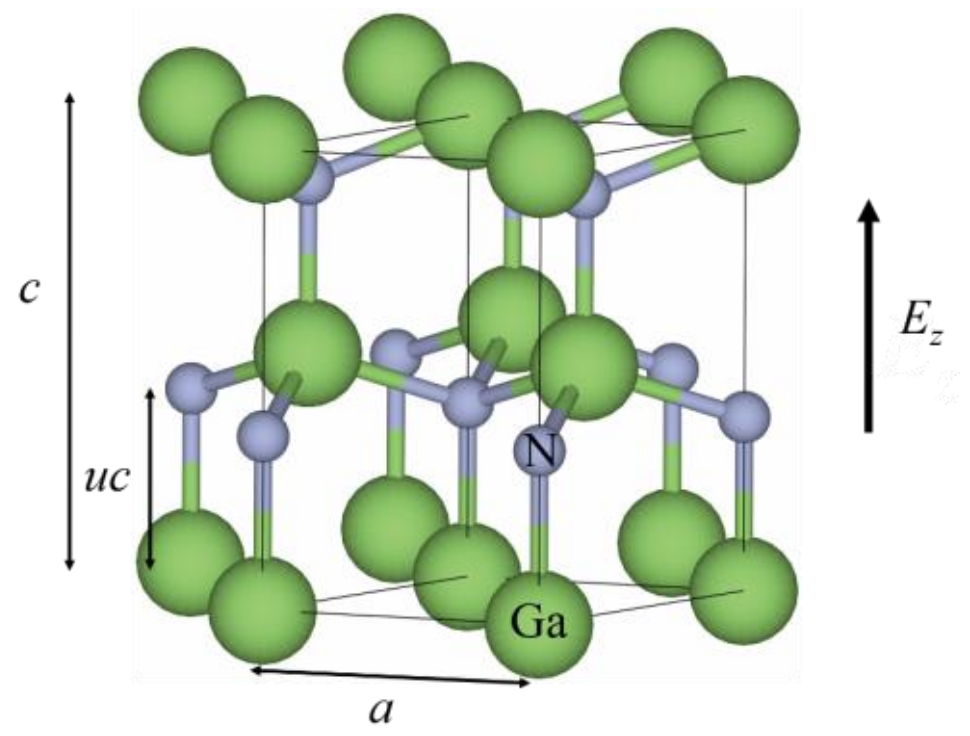

Figure 7. The lattice structure of wurtzite GaN with lattice constants *a* and *c*, and internal parameter *u*. The arrow represents the direction of the positive electric field.

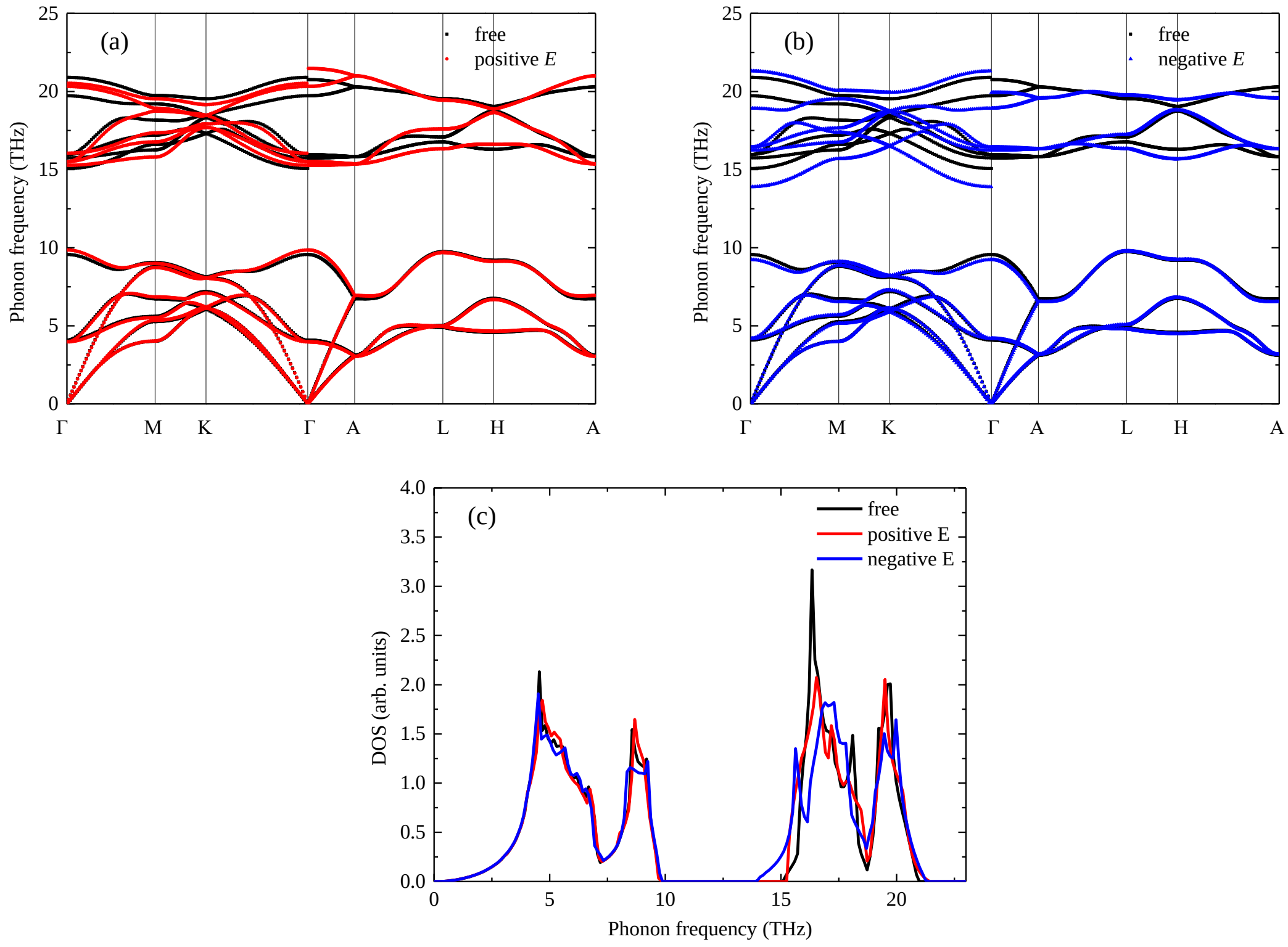

Figure 8. Phonon dispersion relations and DOS of wurtzite GaN with and without lattice deformation from the electric field. The caption "free" in the figures means a condition with zero electric field. And the captions "positive *E*" and "negative *E*" in the figures represent cases with reduced electric fields 0.02 and -0.02, respectively.

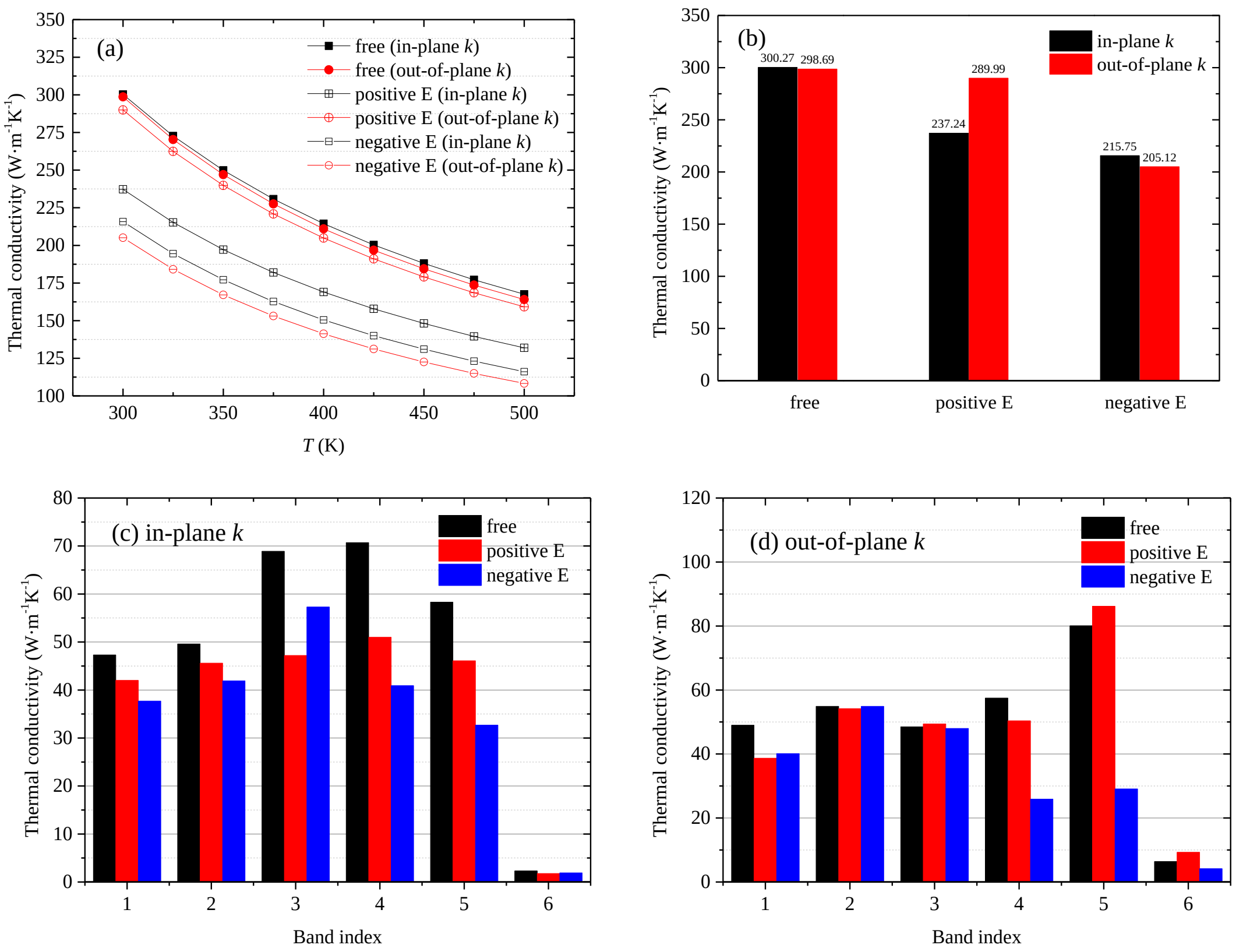

Figure 9. Lattice thermal conductivity of wurtzite GaN with and without lattice deformation from electric field (a) to temperature (b) at room temperature. And lattice thermal conductivity at room temperature for each low-frequency phonon band (bands 1-6) with and without the electric field for (c) in-plane and (d) out-of-plane cases.

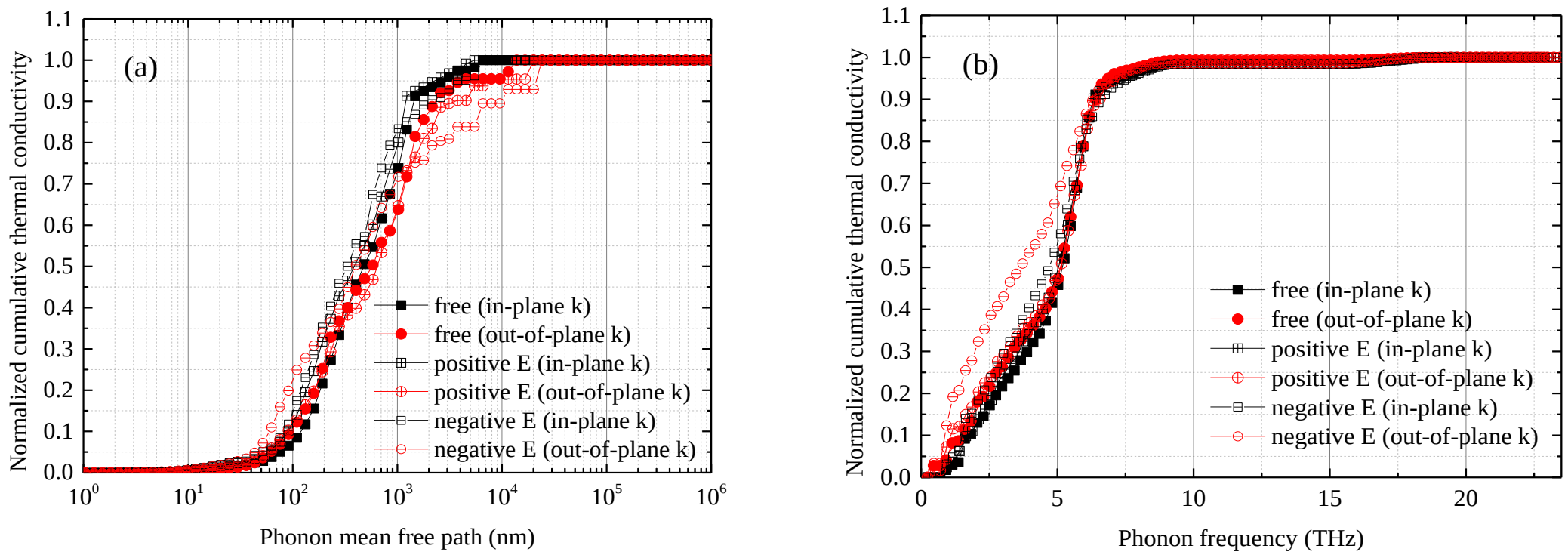

Figure 10. Normalized cumulative lattice thermal conductivity of wurtzite GaN to (a) phonon MFP and (b) phonon frequency, with and without lattice deformation from the electric field at room temperature.

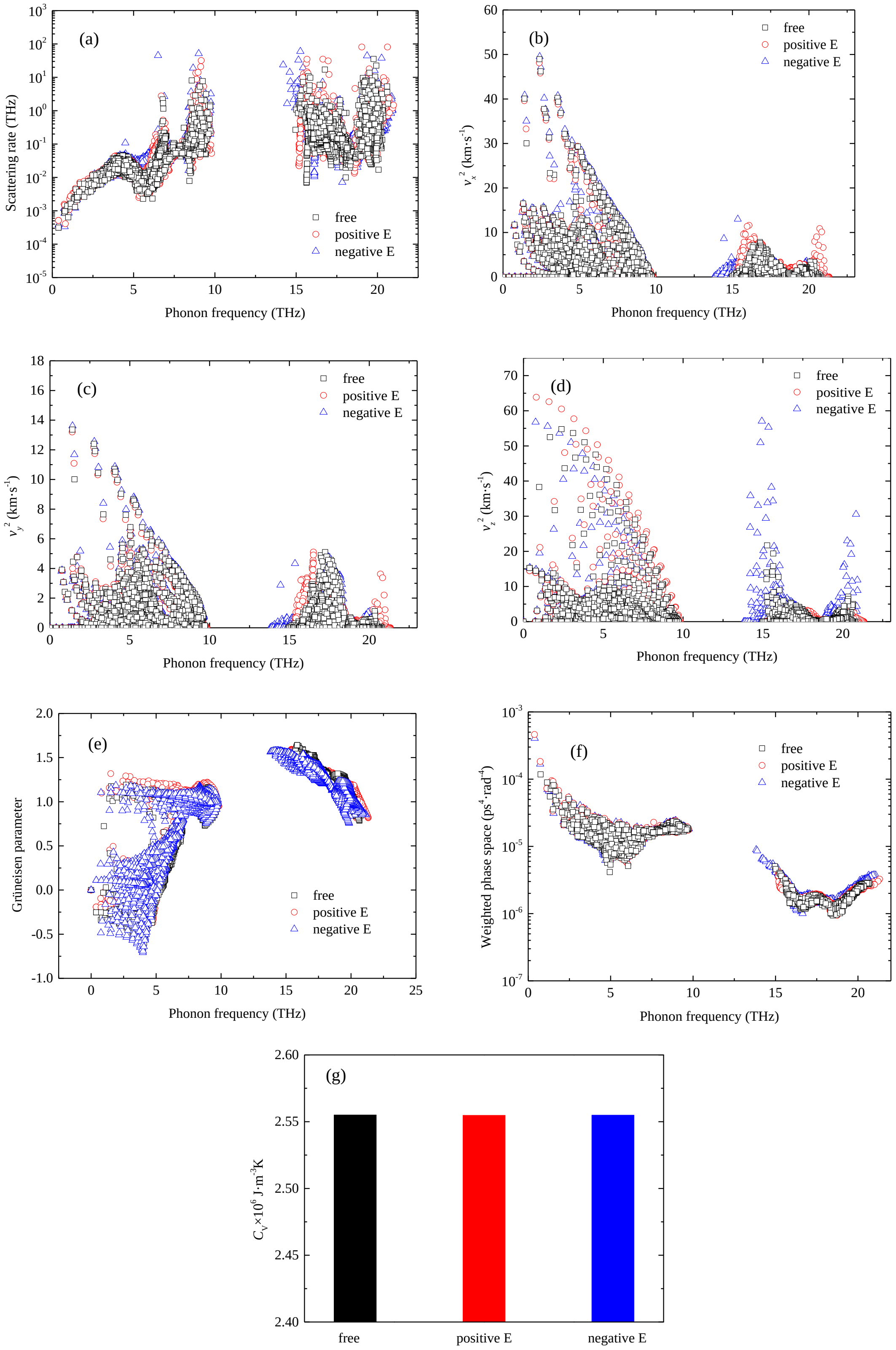

Figure 11. Phonon and thermal properties of wurtzite GaN at room temperature with and without lattice deformation from electric field (a) scattering rate (b) square of the *x* component of group velocity (c) square of the *y* component of group velocity (d) square of the *z* component

of group velocity (e) Grüneisen parameter (f) weighted space group for three phonon process (g) specific heat.

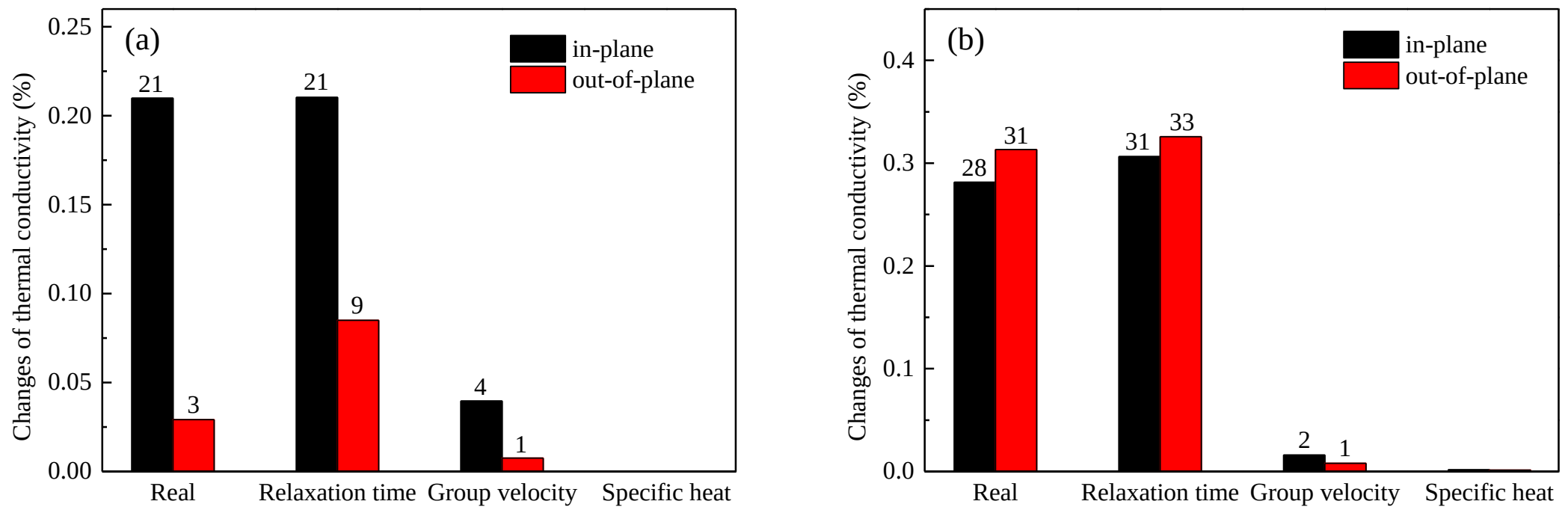

Figure 12. Changes in lattice thermal conductivity (absolute values) at restricted conditions with (a) positive electric field (b) negative electric field.